\documentclass[twocolumn,aps,prl,superscriptaddress]{revtex4}

\usepackage{sidecap}
\usepackage{ulem}
\usepackage{epsfig}
\usepackage{amsmath,amssymb,amsthm}
\usepackage{graphicx}
\usepackage{bm}
\usepackage{color,soul}

\DeclareMathAlphabet{\mathpzc}{OT1}{pzc}{m}{it}

\begin{document}

\renewcommand{\textfraction}{0.00}

% Useful macros:

\newcommand{\vAi}{{\cal A}_{i_1\cdots i_n}}
\newcommand{\vAim}{{\cal A}_{i_1\cdots i_{n-1}}}
\newcommand{\vAbi}{\bar{\cal A}^{i_1\cdots i_n}}
\newcommand{\vAbim}{\bar{\cal A}^{i_1\cdots i_{n-1}}}
\newcommand{\htS}{\hat{S}}
\newcommand{\htR}{\hat{R}}
\newcommand{\htB}{\hat{B}}
\newcommand{\htD}{\hat{D}}
\newcommand{\htV}{\hat{V}}
\newcommand{\cT}{{\cal T}}
\newcommand{\cM}{{\cal M}}
\newcommand{\cMs}{{\cal M}^*}
\newcommand{\vk}{\vec{\mathbf{k}}}
\newcommand{\bk}{\bm{k}}
\newcommand{\kt}{\bm{k}_\perp}
\newcommand{\kp}{k_\perp}
\newcommand{\km}{k_\mathrm{max}}
\newcommand{\vl}{\vec{\mathbf{l}}}
\newcommand{\bl}{\bm{l}}
\newcommand{\bK}{\bm{K}}
\newcommand{\bb}{\bm{b}}
\newcommand{\qm}{q_\mathrm{max}}
\newcommand{\vp}{\vec{\mathbf{p}}}
\newcommand{\bp}{\bm{p}}
\newcommand{\vq}{\vec{\mathbf{q}}}
\newcommand{\bq}{\bm{q}}
\newcommand{\qt}{\bm{q}_\perp}
\newcommand{\qp}{q_\perp}
\newcommand{\bQ}{\bm{Q}}
\newcommand{\vx}{\vec{\mathbf{x}}}
\newcommand{\bx}{\bm{x}}
\newcommand{\tr}{{{\rm Tr\,}}}
\newcommand{\bc}{\textcolor{blue}}

\newcommand{\beq}{\begin{equation}}
\newcommand{\eeq}[1]{\label{#1} \end{equation}}
\newcommand{\ee}{\end{equation}}
\newcommand{\bea}{\begin{eqnarray}}
\newcommand{\eea}{\end{eqnarray}}
\newcommand{\beqar}{\begin{eqnarray}}
\newcommand{\eeqar}[1]{\label{#1}\end{eqnarray}}

\newcommand{\half}{{\textstyle\frac{1}{2}}}
\newcommand{\ben}{\begin{enumerate}}
\newcommand{\een}{\end{enumerate}}
\newcommand{\bit}{\begin{itemize}}
\newcommand{\eit}{\end{itemize}}
\newcommand{\ec}{\end{center}}
\newcommand{\bra}[1]{\langle {#1}|}
\newcommand{\ket}[1]{|{#1}\rangle}
\newcommand{\norm}[2]{\langle{#1}|{#2}\rangle}
\newcommand{\brac}[3]{\langle{#1}|{#2}|{#3}\rangle}
\newcommand{\hilb}{{\cal H}}
\newcommand{\pleft}{\stackrel{\leftarrow}{\partial}}
\newcommand{\pright}{\stackrel{\rightarrow}{\partial}}
%%%%%%%%%%%%%%%%%%%%%%%%%%%%%%%%%%%%%%%%%%%%%%%%%%%%%%%%%%%%%%%%%%%%%%%%%%%%%%

\title{Mass tomography at different momentum ranges in Quark-Gluon Plasma}

\author{Magdalena Djordjevic}
\affiliation{Institute of Physics Belgrade, University of Belgrade, Serbia}

\author{Bojana Blagojevic}
\affiliation{Institute of Physics Belgrade, University of Belgrade, Serbia}

\author{Lidija Zivkovic}
\affiliation{Institute of Physics Belgrade, University of Belgrade, Serbia}

\begin{abstract}
We here show that at lower momentum (i.e. $p_\perp \sim 10$~GeV) single particle suppression for different types of probes exhibit a clear mass hierarchy, which is a direct consequence of the differences in the energy loss, rather than the differences in the initial distributions. On the other hand, we predict that the mass hierarchy is not expected at high momentum (i.e. $p_\perp \sim 100$~GeV); i.e. while we surprisingly predict that suppression for charged hadrons will be somewhat {\it smaller} than the suppression for heavy mesons, we find that this difference will be a consequence of fragmentation functions, not the finite mass effects. That is, apart from the fragmentation functions,  the probes of different masses exhibit nearly the same suppression in the high momentum region. We also argue that the same insensitivity on the probe types also appears for jets. In particular, the experimental data in the momentum regions where they exist for both types of probes, show similar suppressions of charged hadrons and inclusive jet data. Interestingly, we also find that our state-of-the-art suppression predictions for high momentum single particles are also in agreement with the jet suppression data, where the reasons behind this agreement yet remain to be understood. Finally, the available jet data also show (though with large error bars) an overlap between b-jets (heavy) and inclusive jets (light) probes. Consequently, our results suggest that single particles in the momentum region below 50 GeV present an excellent tool for mass tomography, while high momentum single particles and (possibly) jets are rather insensitive to the details of the interaction with Quark-Gluon Plasma.
\end{abstract}

%\pacs{12.38.Mh; 24.85.+p; 25.75.-q}

\maketitle

\section{Introduction}

Quark-gluon plasma (QGP) is a new state of matter~\cite{Collins,Baym} consisting of interacting quarks, antiquarks and gluons. Such new state of matter is created in ultra-relativistic heavy ion collisions at Relativistic Heavy Ion Collider (RHIC) and Large Hadron Collider (LHC). Rare high momentum probes, which are created in such collisions and which transverse QGP, are excellent probes of this extreme form of matter~\cite{QGP1,QGP2,QGP3}. As these probes have different masses and consequently interact with the medium in a different manner, such mass tomography allows investigating properties of the interactions with the medium~\cite{Bjorken,DG_PRL,Kharzeev}. Furthermore, as higher momentum  ranges become increasingly available at the LHC experiments, there are both different probes and a wide range of their momentum, which become available for such mass tomography. However, there is now a question which exactly probes, and momentum ranges, are optimal for such tomography, i.e. will lead to different behavior that can provide new information about interactions with the medium. To address this question, we will in this paper concentrate on the nuclear modification factor ($R_{AA}$), as suppression is traditionally considered to be an excellent observable for mass tomography.

As an example, it was previously widely expected that such clear distinction between the suppression patterns will be provided by the measurements of charged hadron (light) and D meson (heavy) probes (see e.g.~\cite{DG_PRL,Kharzeev}). However, as shown by both the experimental data~\cite{ALICE_D,ALICE_h} and theoretical predictions~\cite{MD_PRL}, these two probes have the same suppression at least in the momentum region between 10 and 50 GeV, which is a consequence of a serendipitous interplay between energy loss and fragmentation functions. Below 10 GeV, there exists a small difference in the $R_{AA}$s between D mesons and charged hadrons; however, this difference in the suppressions is both small and somewhat influenced by the fragmentation functions~\cite{MD_PRL}, so it is, unfortunately, not suitable for extracting any reliable conclusions. Furthermore, at high momentum, recent jet measurements indicate (though with large error bars) the same suppression for b-jets~\cite{CMS_BJets}, and inclusive (light) jets~\cite{CMS_Jets,ATLAS_Jets}. Consequently, there is a nontrivial question of what exactly probes and momentum ranges can be used for obtaining new information on probe-medium interactions. Addressing this will, in turn, allow optimally exploiting experimental efforts and provide further tests of our understanding of QCD matter. Systematically testing the mass tomography effects, for different probes, and at wide momentum ranges, will be the main goal of this paper.

To achieve this goal, we will here use our state-of-the-art dynamical energy loss formalism~\cite{MD_PRC,DH_PRL}, which removes a widely used static approximation and takes into account interactions of the probe with the moving (dynamical) medium constituents. Through this, it consistently treats both radiative~\cite{MD_PRC,DH_PRL} and collisional~\cite{MD_Coll} energy loss, which has been shown to be crucial for quantitatively and qualitatively explaining the experimental data~\cite{BD_JPG}. Additionally, the formalism also takes into account finite magnetic mass~\cite{MD_MagnMass} and running coupling~\cite{MD_PLB}, and is integrated in an up-to-date numerical procedure, which includes path-length~\cite{WHDG} and multi-gluon~\cite{GLV_suppress} fluctuations. The formalism was previously shown to be consistent with the wide range of suppression data corresponding to different probes and experimental conditions~\cite{MD_PLB,MD_PRL,DDB_PLB}. Importantly, {\it no free parameters} are used in comparing predictions with the experimental data. The same parameter set, corresponding to the standard literature values, will be used in this paper, so that the generated predictions will be also constrained by an agreement with a wealth of previous data.

We will here generate single particle $R_{AA}$ predictions at both lower momentum (i.e. $p_\perp \sim 10$~GeV) and high momentum (i.e. $p_\perp \sim 100$~GeV) regions. Our predictions are applicable for both 2.76 TeV and 5.02 TeV collision energies, since we here predict the same suppression at these two collision energies for light flavor, while we previously~\cite{MD_5TeV} predicted the same suppression at these energies for heavy flavor. Comparing these predictions with single particle $R_{AA}$ measurements will allow investigating how suppression depends on the mass hierarchy in different momentum regions, particularly since high precision single particle $R_{AA}$ measurements are (or will soon become) available at both lower and high momentum ranges. In the high momentum range, we will generate predictions for 5.02 TeV collision energy, where preliminary experimental data are currently becoming available. The high momentum single particle predictions are not available for 2.76 TeV, so in this range, we will compare our single particle predictions for 5.02 TeV (which are also applicable to 2.76 TeV, see above) with the available jet measurements. The comparison of the single particle predictions with the available jet data is motivated by the fact that, in the momentum region where both (limited) single particle and jet $R_{AA}$ data exist, these two observables show the same suppression within the errorbars, as we present below. This observation leads to a question of how the leading particle $R_{AA}$ predictions, done with state-of-the-art dynamical energy loss model, compare with the whole jet $R_{AA}$, which we will here address. Consequently, we will here provide a systematic investigation of how the predicted suppression depends on the probe type, the momentum and collision energy range, and how these predictions compare with various available data.

\section{Methods}
The numerical framework for generating suppression predictions is presented in detail in~\cite{MD_PLB}. We below briefly outline the main steps in this procedure.

We study the angular averaged nuclear modification factor $R_{AA}$, which is established as an excellent probe for studying the interaction of high-momentum particles with QGP. $R_{AA}$ is the ratio of the quenched spectrum in $A+A$ collisions to the spectrum in $p+p$ collisions, scaled by the number of binary collisions $N_{\mathrm{bin}}$:
\begin{eqnarray}
R_{AA}(p_\perp)=\dfrac{dN_{AA}/dp_\perp}{N_{\mathrm{bin}} dN_{pp}/dp_\perp}
\label{RAA}
\end{eqnarray}

To calculate the quenched spectra of light and heavy probes, we use the generic pQCD convolution, which consists of the following steps:
\begin{eqnarray}
\frac{E_f d^3\sigma}{dp_{\perp,f}^3} &=& \frac{E_i d^3\sigma(Q)}{dp^3_{\perp,i}}
 \otimes
{P(E_i \rightarrow E_f )} \otimes \\
&&\otimes D(Q \to H_Q) \otimes f(H_Q \to J/\psi). \;
\label{schem} \end{eqnarray}
Here "i"  and "f" subscripts correspond, respectively, to "initial" and "final", $E$ is energy, $p_\perp$ is momentum, $Q$ denotes partons (quarks and gluons), and the terms in the equation correspond to the following:
\begin{itemize}

\item[{\it i)}] $E_i d^3\sigma(Q)/dp_{\perp,i}^3$ denotes the initial parton spectrum. For light quarks and gluons, the spectrum is extracted from~\cite{Vitev0912}, while for charm and bottom quarks, the spectrum is extracted from~\cite{Cacciari:2012}.
\item[{\it ii)}]  $P(E_i \rightarrow E_f )$ is the energy loss probability. The probability is generalized to include both collisional~\cite{MD_Coll} and radiative~\cite{MD_PRC,DH_PRL} energy loss in the same framework (i.e. realistic finite size dynamical QCD medium), which abolishes the widely used approximation of static scattering centers. It is also recently improved to include path-length~\cite{WHDG} and multi-gluon~\cite{GLV_suppress} fluctuations, as well as running coupling~\cite{MD_PLB} and finite magnetic mass~\cite{MD_MagnMass}.
\item[{\it iii)}] $D(Q \to H_Q)$ is the parton to hadron $H_Q$ fragmentation function. For light hadrons, D and B mesons we use DSS~\cite{DSS}, BCFY~\cite{BCFY} and KLP~\cite{KLP} fragmentation functions, respectively. Note, however, that for heavy flavor, fragmentation functions do not influence the suppression of heavy mesons~\cite{MD_PRL}. That is, heavy meson $R_{AA}$ is a true probe of heavy quark $R_{AA}$. \smallskip
\item[{\it iv)}] For non-prompt $J/\Psi$, we also have to include the decay of B meson to $J/\psi$, which is represented by the function $f(H_Q \to J/\psi)$ and obtained according to~\cite{Cacciari:2012}.\end{itemize}
\medskip

To generate the suppression predictions for light and heavy flavor observables in Pb+Pb collisions at the LHC experiments, we used the following parameter set: QGP with perturbative QCD scale of $\Lambda_{QCD}=0.2$~GeV and effective light quark flavors $n_f{\,=\,}3$. In the calculations, as a starting point we use an effective temperature of 304 MeV for 0-40$\%$ centrality Pb+Pb collisions at the LHC~\cite{ALICE_T} experiments (as extracted by ALICE); the average temperature for every centrality region is then determined according the procedure given in~\cite{DDB_PLB}. Also, for every centrality region, we use different path-length distributions, which are provided to us by~\cite{Dainese}. The light quark mass is assumed to be dominated by the thermal mass $M{\,=\,}\mu_E/\sqrt{6}$, where temperature dependent Debye mass
$\mu_E$ is obtained from~\cite{Peshier}. The gluon mass is $m_g=\mu_E/\sqrt{2}$~\cite{DG_TM}, while the charm and the bottom masses are, $M{\,=\,}1.2$\,GeV  and $M{\,=\,}4.75$\,GeV, respectively. Magnetic to electric mass ratio is $0.4 < \mu_M/\mu_E < 0.6$, as extracted from several independent non-perturbative QCD calculations~\cite{Maezawa,Nakamura,Hart,Bak}, so the uncertainty in the predictions, presented in the next section, will come from this range of screening masses ratio. Note that our model uses no free parameters in comparison with the experimental data, that is all the parameters correspond to standard literature values.

\section{Numerical results}

In this section, we will generate predictions which correspond to the probe suppression at both lower ($\sim 10$ GeV) and high ($\sim 100$ GeV) momentum ranges. At high momentum ranges, we will also compare the single particle and jet measurements with each other, and with the generated theoretical predictions. The predictions will be generated both as a function of probe momentum and the number of participants and for both light and heavy flavor observables.

%%%%%%%%%%%%%%%%%%%%%%%% Fig. 1 %%%%%%%%%%%%%%%%%%%%%%%%%%%%%%%%%%%%%%%%%%%%
\begin{figure*}
\epsfig{file=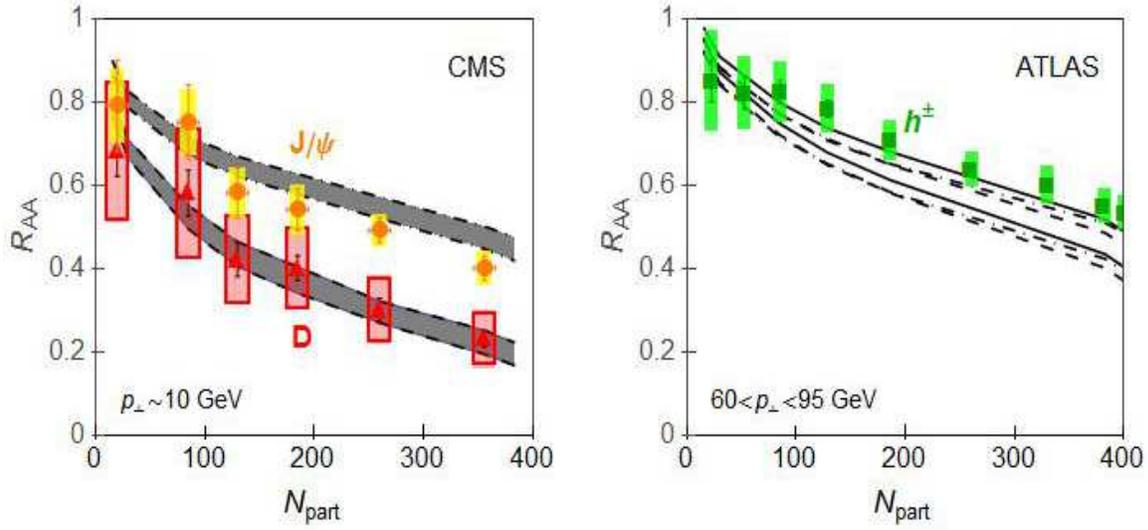,width=6in,height=3in,clip=5,angle=0}
\vspace*{-0.4cm}
\caption{{\bf $R_{AA}$ {\it vs.} $N_{part}$ for single particles at the 2.76 TeV Pb+Pb collisions at the LHC experiments.} {\it Left panel:} Theoretical predictions for $R_{AA}$ {\it vs.} $N_{part}$ are compared with CMS experimental data for D mesons~\cite{CMS_D} (red triangles) and non-prompt $J/\Psi$~\cite{CMS_JPsi} (orange circles) in, respectively, $8<p_\perp<16$ GeV and $6.5<p_\perp<30$ GeV momentum regions. Gray bands with dashed, and dot-dashed boundaries, respectively, correspond to predictions for D mesons and non-prompt $J/\Psi$ in corresponding momentum regions.  {\it Right panel:} Theoretical predictions for $h^\pm$, D and B meson $R_{AA}$ {\it vs.} $N_{part}$ in $60<p_\perp<95$ GeV momentum region are, respectively, provided as white bands with full, dashed and dot-dashed boundaries. The $h^\pm$ predictions are compared with ATLAS (green squares)~\cite{ATLAS_CH} $h^\pm$ experimental data in the same  momentum region.  On each panel, the upper (lower) boundary of each gray (or  white) band corresponds to $\mu_M/\mu_E =0.6$ ($\mu_M/\mu_E =0.4$). }
\label{Fig1}
\end{figure*}
%%%%%%%%%%%%%%%

%%%%%%%%%%%%%%%%%%%%%%%%%%%%%%%%%%%%%%%%%%%%%%%%%%%%%%%%%%%%%

We first show predictions for the suppression dependence on the number of participants at 2.76 TeV Pb+Pb collision energy. In Fig.~1 (left) we compare predictions with the data in the lower momentum range ($p_\perp\sim 10$~GeV), while in Fig.~1 (right) we compare predictions with the data in the high momentum range ($p_\perp\sim 100$~GeV). Note that the formalism is developed under the assumption that $M^2/E^2 \ll 1$, so, for all types of quarks (both light and heavy), our predictions are valid for $p_\perp \gtrsim 10$ GeV. The predictions in Fig.~1 (left) are generated for $J/\Psi$ and D mesons, and compared with the corresponding CMS experimental data~\cite{CMS_D} - D meson data from ALICE~\cite{ALICE_D}, not shown for figure representation, are consistent with CMS D meson data. Also, the charged hadrons (light probes) are not shown in Fig.~1 (left) for clarity, as it was previously shown that both experimental data~\cite{ALICE_D,ALICE_h} and theoretical predictions~\cite{MD_PRL} largely overlap with those for D mesons. Since charged hadrons are indirect/complex probes, composed of both light quarks and gluons with nontrivial effect of fragmentation functions~\cite{MD_PRL}, for mass tomography it is clearly beneficial to, whenever possible, concentrate on D mesons (clear charm quark probes~\cite{MD_PRL}) instead of charged hadrons. In Fig.~1 (right), the theoretical predictions for charged hadrons, D and B mesons are generated and shown together with the ATLAS charged hadron experimental data.

Clear distinction in predictions between lower and high $p_\perp$ ranges are observed. In addition, for lower $p_\perp$ (the left panel of Fig. 1), it is obvious that the light and heavy flavor suppressions are significantly different. On the other hand, in the high $p_\perp \sim 100$ GeV range (the right panel of Fig. 1), the predictions for all the probes (both light and heavy) almost overlap with each other. From pQCD perspective, a reason for similar suppressions at high momentum is that the mass of the probe becomes small compared to its momentum, so the relevance of mass effects should also become small in this region. However, while plausible/expected from pQCD perspective, this prediction can be quite distinct in other approaches, as e.g. AdS-CFT predicts a clear suppression mass hierarchy, even for high momentum regions~\cite{Horowitz1,Horowitz2}.

The experimental data shown in Fig. 1 are in a good agreement with the generated theoretical predictions. Moreover, these data also confirm the predicted qualitative distinction between the light/charm and bottom suppressions at lower momentum region. At the higher momentum range, such comparison between the light and heavy flavor experimental data cannot be made, as the corresponding single particle data are currently available only for charged hadrons. Therefore, the overlap of the light and heavy flavor suppressions at high momentum ranges, provides a clear prediction to be tested by the upcoming experiments.

%%%%%%%%%%%%%%%%%%%%%%%% Fig. 2 %%%%%%%%%%%%%%%%%%%%%%%%%%%%%%%%%%%%%%%%%%%%
\begin{figure*}
\epsfig{file=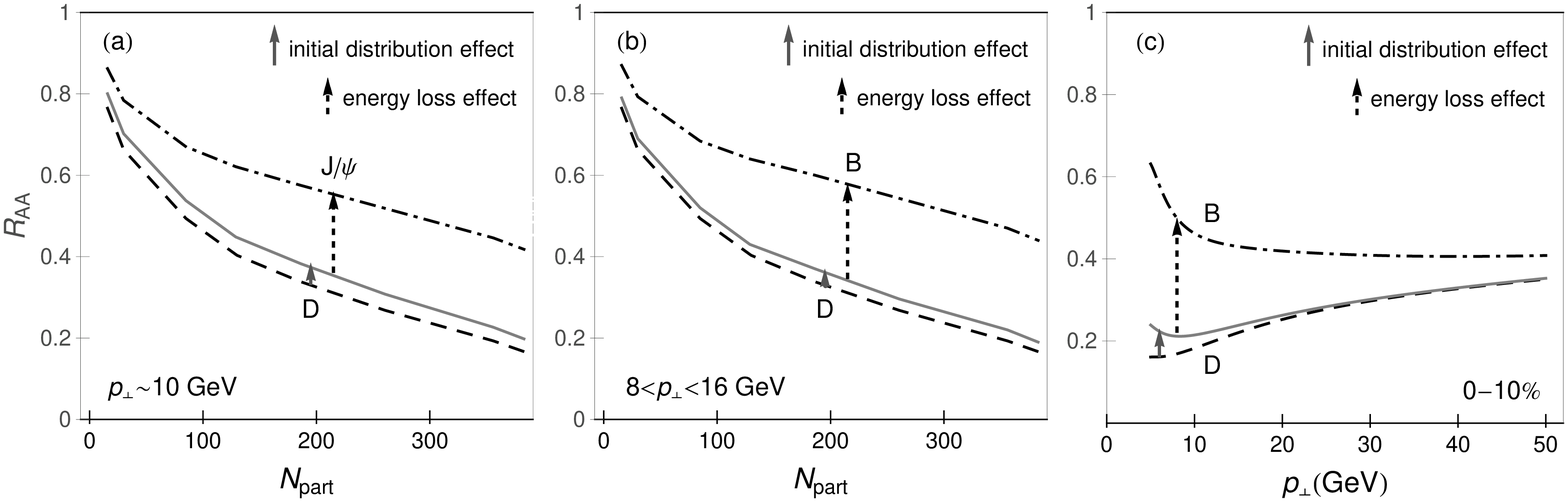,width=6.5in,height=2.3in,clip=5,angle=0}%\end{minipage}
\vspace*{-0.4cm}
\caption{{\bf Suppression contributions.} {\it Left panel} (a): Theoretical predictions for $R_{AA}$ {\it vs.} $N_{part}$ are compared for D mesons (dashed curve, $8<p_\perp<16$ GeV momentum region) and non-prompt $J/\Psi$ (dot-dashed curve, $6.5<p_\perp<30$ GeV momentum region). Gray curve shows the analogous non-prompt $J/\Psi$ predictions, if the originating bottom quark would have the same energy loss as charm quark in QGP. {\it Middle panel} (b): Theoretical predictions for $R_{AA}$ {\it vs.} $N_{part}$ are compared for D and B mesons in $8<p_\perp<16$ GeV momentum region. Comparing the left and the middle panels shows the effect of decay functions to the contributions analyzed in the left panel.  {\it Right panel} (c): Theoretical predictions for $R_{AA}$ {\it vs.} $p_\perp$ are compared for D  and B mesons. In the middle and the right panels, the curve legend is the same as in the left panel. On each panel, the full arrow points to the contribution of the different initial distributions to the difference in the suppression between D meson and non-prompt $J/\Psi$ (or B meson), while the dashed arrow points to the contribution of the different energy losses to the difference between D meson and the non-prompt $J/\Psi$ (or B meson) suppression.  Magnetic to electric mass ratio is set to $\mu_M/\mu_E =0.4$. }
\label{Fig2}
%\end{minipage}
\end{figure*}
%%%%%%%%%%%%%%%%%%%%%%%%%%%%%%%%%%%%%%%%%%%%%%%%%%%%%%%%%%%%%%%%%%%%%%%%%%%%

For understanding the difference between D meson and non-prompt $J/\Psi$ suppressions, we studied the importance of different contributions to this suppression difference. Regarding this, note that it has been considered that this difference may largely originate from the differences in the initial distributions between charm and bottom quarks, rather than the difference in their energy losses~\cite{ALICE_2015}. We show in Fig. 2 (left) that this is not the case, i.e. the contribution to the suppression difference from the initial distributions is small, while the contribution due to the different energy loss is substantially larger. Note however that the contributions shown in the Fig. 2 (left) are not pure effects of initial distribution and energy loss. This is since $J/\Psi$ suppression is not a direct probe of b quarks, i.e. it includes a decay from B mesons to $J/\Psi$, i.e. $f(B \rightarrow J/\Psi)$. Consequently, to exclude the decay contribution from these two effects, in Fig. 2 (middle) we show the same contributions but with the B mesons (clear b quark probe) instead of $J/\Psi$. We see that, in the case of B mesons, the energy loss contribution becomes even larger, while the initial distribution becomes even smaller. Therefore, the strong mass dependence of the suppression, which is observed and predicted at lower momentum ranges, is clearly a consequence of the differences in the energy loss, rather than the consequence of the initial distributions or decay. Furthermore, we show in Fig. 2 (right) that there is no momentum region in which initial distribution makes a significant effect on the suppression difference between different types of probes. Therefore, studying the difference between D and B meson suppression patterns in lower momentum region is not influenced by the production, fragmentation and the decay, and therefore allows directly assessing how different probes interact with QGP.

%%%%%%%%%%%%%%%%%%%%%%%% Fig. 4 %%%%%%%%%%%%%%%%%%%%%%%%%%%%%%%%%%%%%%%%%%%%
\begin{figure*}
\epsfig{file=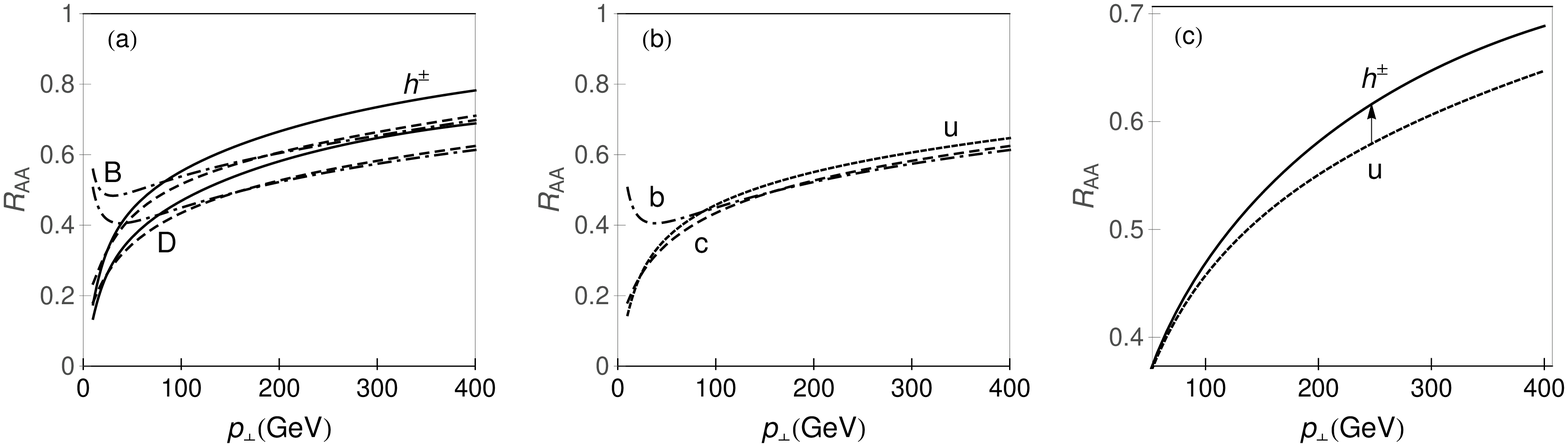,width=6.7in,height=2.3in,clip=5,angle=0}%\end{minipage}
\vspace*{-0.4cm}
\caption{{\bf $R_{AA}$ {\it vs.} $p_\perp$ for single particles at the 5.02 TeV Pb+Pb 0-10\% central collisions at the LHC.} {\it Left panel} (a): Theoretical predictions for $h^\pm$, D and B meson $R_{AA}$ {\it vs.} $p_\perp$ are, respectively, given as white bands with full, dashed and dot-dashed boundaries. The upper (lower) boundary of each band corresponds to $\mu_M/\mu_E =0.6$ ($\mu_M/\mu_E =0.4$). {\it Middle panel} (b): Theoretical predictions for bare quark $R_{AA}$ {\it vs.} $p_\perp$ are shown for $u$ (dotted curve), $c$ (dashed curve) and $b$ (dot-dashed curve). $\mu_M/\mu_E$ ratio is set to 0.4. {\it Right panel} (c):  Theoretical predictions for $R_{AA}$ {\it vs.} $p_\perp$ are compared for $u$ (dotted curve) with $h^\pm$ (full curve). $\mu_M/\mu_E$ ratio is set to 0.4.}
\label{RAA5TeV}
%\end{minipage}
\end{figure*}
%%%%%%%%%%%%%%%%%%%%%%%%%%
%%%%%%%%%%%%%%%%%%%%%%%%%%%%%%%%%%%%%%%%%%%%%%%%%%

While there are currently only limited data for single particles at high momentum, these type of measurements are expected to become increasingly available at 5.02 TeV Pb+Pb collisions at the LHC. At 5.02 TeV collision energy, the $R_{AA}$ measurements for charged hadrons are expected to become available up to $p_\perp \sim 400$ GeV, for D mesons the $R_{AA}$ measurements might be available up to $p_\perp \sim 200$ GeV, while for bottom mesons the measurements are expected up to $p_\perp \sim 100$ GeV~\cite{YenJieLee} and possibly even higher. It is, therefore, useful providing single particle $R_{AA}$ predictions in the high momentum region, and studying the effects of high $p_\perp$ mass tomography.

With this goal, in Fig.~\ref{RAA5TeV}, we provide predictions for charged hadrons, D and B mesons $R_{AA}$ at 5.02 TeV 0-10\% central Pb+Pb collisions at the LHC. From the right panel, we see that at $p_\perp \sim 100$ GeV, all types of probes show similar suppressions, as supported by the right panel of Fig. 1. However, for $p_\perp > 100$ GeV, we also observe that, while D and B (i.e. heavy) meson $R_{AA}$s become almost identical, the $h^\pm$ (i.e. light hardon) $R_{AA}$ shows a surprising tendency for a {\it lower} suppression compared to heavy mesons. Moreover, we see that the difference between light and heavy meson suppression increase with increasing $p_\perp$, leading to more than $10\%$ higher $R_{AA}$ for $h^\pm$ compared to heavy mesons at $p_\perp > 150$ GeV. That is our observation is that, contrary to the 2.76 TeV collision energy where overlap between $h^\pm$ and D meson $R_{AA}$ was observed for the entire momentum region where both data are available ($p_\perp<50$ GeV), we here predict that increasing momentum (above $p_\perp$ of 100 GeV) will lead to the separation in the $R_{AA}$ of these two observables, but in a different direction than intuitively expected.

A naive conclusion from this prediction would be that, for highly energetic partons, the light partons start to lose notably less energy compared to heavy partons, which is not in accordance with pQCD, as discussed just after the Fig. 1. To further investigate this issue, in the middle panel of Fig.~\ref{RAA5TeV}, we compare $R_{AA}$ predictions for bare quarks, i.e. for up, charm and bottom quarks. We here observe that for $p_\perp >100$ GeV, and in accordance with pQCD, finite mass effects for all types of quarks become negligible, leading to the same suppressions for both light and heavy flavor partons. However, from the right panel of Fig.~\ref{RAA5TeV}, we see that the nonintuitive result observed in the left panel of Fig.~\ref{RAA5TeV}, is a consequence of fragmentation function effect on the light partons that compose the charged hadrons. That is, the effect of fragmentation functions on the light quarks is to decrease their suppression (noted by the vertical arrow in the right panel of Fig.~3); the gluon contribution (partially) compensate this effect (as discussed in~\cite{MD_PRL}), but for $p_\perp >100$ GeV, the gluon contribution, and therefore the gluon compensation effect, is small. Due to this, we conclude that, if our predicted larger $R_{AA}$ for $h^\pm$ compared to heavy flavor (D and B) in the high $p_\perp$ region is indeed experimentally observed, this increase will be a pure consequence of the fragmentation function effect, and therefore not related with the mass tomography in the QGP.

Moreover, the predictions presented in Fig.~\ref{RAA5TeV}  show that the mass tomography effects can be clearly observed below 50 GeV. In particular, we see that below 50 GeV, bottom suppression significantly differs compared to charm and light probes. On the other hand, such a distinction does not appear for high probe momentum (above 50 GeV) where all the suppression predictions nearly overlap (apart from fragmentation functions effect discussed above). Furthermore, we also showed that indirect bottom probes (i.e. non-prompt $J/\Psi$) lower the dead-cone~\cite{Kharzeev} effect compared to the clear B meson probes. Consequently, we propose that one should concentrate on the lower momentum range and on the difference between the B meson suppression on one side, and D meson/charged hadron suppression on the other side, for observing significant mass tomography effects.
%%%%%%%%%%%%%%%%%%%%%%%% Fig. 4 %%%%%%%%%%%%%%%%%%%%%%%%%%%%%%%%%%%%%%%%%%%%
\begin{figure}
\epsfig{file=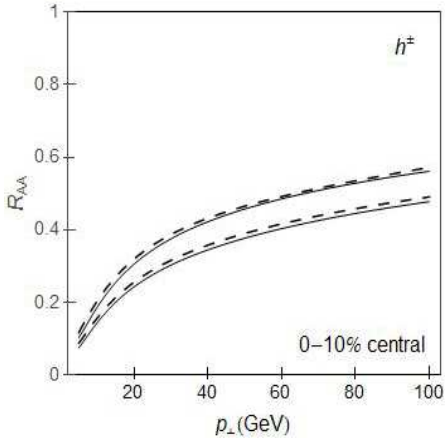,width=2.3in,height=2.3in,clip=5,angle=0}%\end{minipage}
\vspace*{-0.4cm}
\caption{\textbf {Comparison of $R_{AA}$ predictions for charge hadrons at 2.76 and 5.02 TeV.} Charged hadron suppression predictions, as a function of transverse momentum, are shown. $R_{AA}$ predictions at 5.02 TeV (2.76 TeV) 0-10\% central Pb+Pb collisions at the LHC.  are presented as white bands with full (dashed) boundaries. The upper (lower) boundary of each band corresponds to $\mu_M/\mu_E =0.6$ ($\mu_M/\mu_E =0.4$).}
\label{HadronRAA}
%\end{minipage}
\end{figure}
%%%%%%%%%%%%%%%%%%%%%%%%%%%%%%%%%%%%%%%%%%%%%%%%%%%%%%%%%%%%%%%%%%%%%%%%%%%%

Furthermore, in~\cite{MD_5TeV}, we have shown that, for heavy flavor, the $R_{AA}$ predictions for 5.02 TeV and 2.76 TeV overlap with each other, due to interplay between energy loss and initial distributions. In Fig.~\ref{HadronRAA}, we show that the same conclusion is valid for charged hadrons as well. We therefore conclude that all the predictions/observations presented in this paper are valid for both 2.76 TeV and 5.02 TeV collision energies.

%%%%%%%%%%%%%%%%%%%%%%%% Fig. 3 %%%%%%%%%%%%%%%%%%%%%%%%%%%%%%%%%%%%%%%%%%%%
\begin{figure*}
\epsfig{file=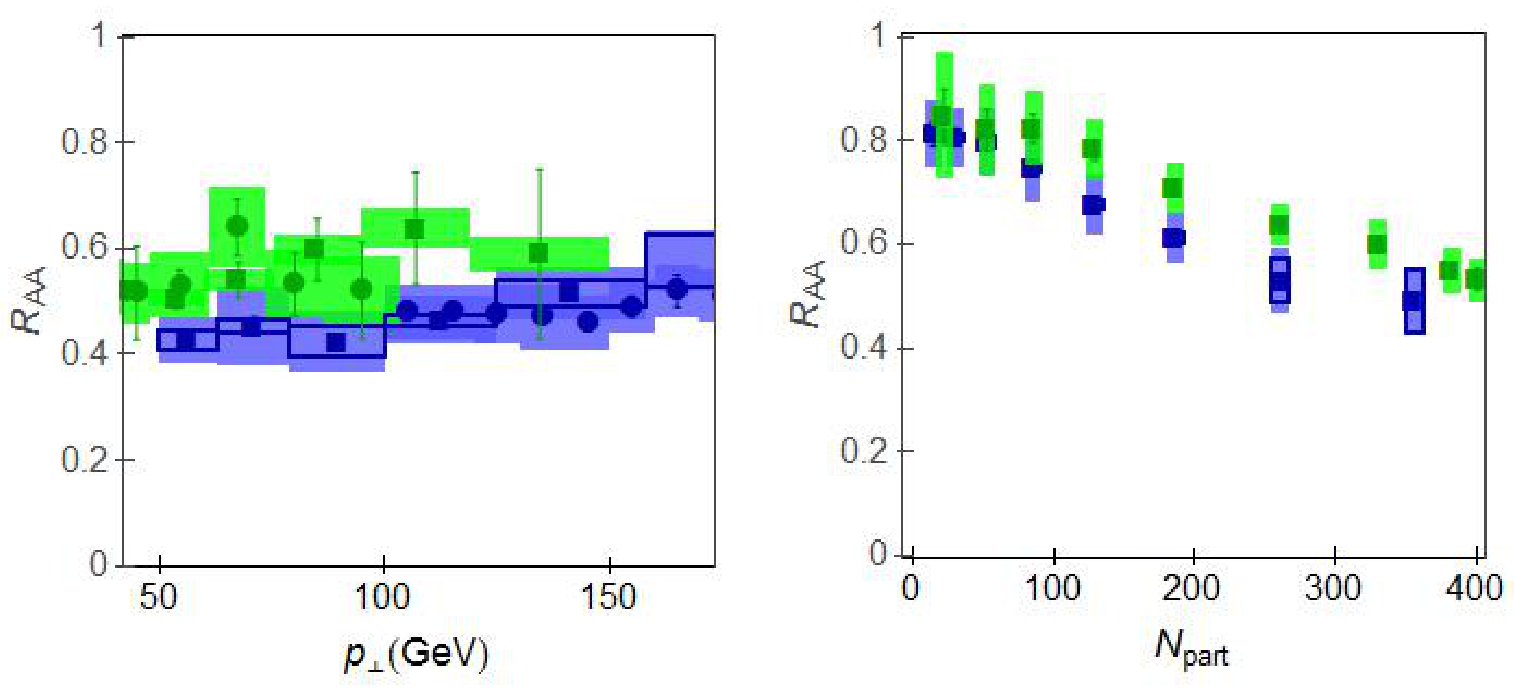,width=6in,height=3in,clip=5,angle=0}%\end{minipage}
\vspace*{-0.4cm}
\caption{{\bf Comparison of single particle and jet suppression data at the LHC experiments.} {\it Left panel:} $R_{AA}$ {\it vs.} $p_\perp$ experimental data are compared for inclusive jets from ATLAS~\cite{ATLAS_Jets} (blue squares) and CMS~\cite{CMS_Jets} (blue circles) and charged hadrons~\cite{ATLAS_CH} from ATLAS (green squares) and CMS~\cite{CMS_CH} (green circles). ATLAS jet data correspond to 0-10\% centrality, while the other data correspond to 0-5\% centrality.  {\it Right panel:} $R_{AA}$ {\it vs.} $N_{part}$ ATLAS experimental data are compared for inclusive jets~\cite{ATLAS_Jets} (blue squares, $63<p_\perp<80$~GeV) and charged hadrons~\cite{ATLAS_CH} (green squares, $65<p_\perp<90$~GeV). }
\label{JetsVsSingle}
%\end{minipage}
\end{figure*}
%%%%%%%%%%%%%%%%%%%%%%%%%%%%%%%%%%%%%%%%%%%%%%%%%%%%%%%%%%%%%%%%%%%%%%%%%%%%

With regard to this, we note that, while high $p_\perp$ data are not available for single particles at the currently available 2.76 TeV collisions, high $p_\perp$ data are abundant for jets. As our theoretical predictions for single particle $R_{AA}$ data at 5.02 TeV are also applicable for 2.76 TeV collision energy, it is tempting to compare these predictions with the available jet data at 2.76 TeV collision energy. Before comparing single particle predictions with the jet data, we address the same comparison with the experimental data, i.e. we start by asking how the single particle data and the jet measurements correspond to each other, in the momentum range where both are available. Consequently, in Fig.~\ref{JetsVsSingle}, we compare the available experimental data for charged hadrons (the green squares and circles) and inclusive jets (the blue squares and circles). In the left panel, we show the comparison of the measured suppression dependencies on the probe momentum (for the similar, fixed centrality region), while in the right panel, we show the comparison of the measured suppression dependencies on the number of participants (for the similar, fixed momentum region). Therefore, one can see that similar suppressions are observed for single particles and jets, i.e. while the inclusive jets show a somewhat higher suppression compared to charged hadrons, they are the same within the error bars.

%%%%%%%%%%%%%%%%%%%%%%%% Fig. 4 %%%%%%%%%%%%%%%%%%%%%%%%%%%%%%%%%%%%%%%%%%%%
\begin{figure*}
\epsfig{file=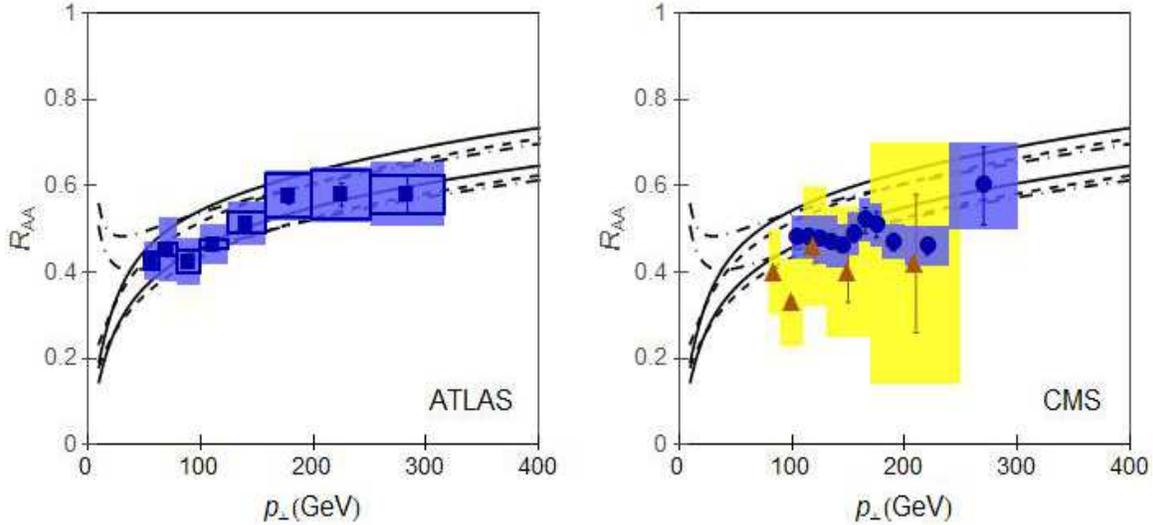,width=6.in,height=3.in,clip=5,angle=0}%\end{minipage}
\vspace*{-0.4cm}
\caption{{\bf Single particle suppression predictions {\it vs.} jet data.} {\it Left panel:} Theoretical $R_{AA}$ {\it vs.} $p_\perp$ predictions for single particles are compared with 0-10\% centrality ATLAS experimental data for inclusive jets~\cite{ATLAS_Jets} (blue squares). {\it Right panel:} $R_{AA}$ {\it vs.} $p_\perp$ single particle predictions are compared with CMS experimental data for inclusive jets~\cite{CMS_Jets} (blue circles, 0-5\% centrality) and b-jets~\cite{CMS_BJets} (orange triangles, 0-10\% centrality). On each panel, white bands with dashed, dot-dashed and full boundaries, respectively, correspond to charm, bottom and light flavor predictions, and the upper (lower) boundary of each band corresponds to $\mu_M/\mu_E =0.6$ ($\mu_M/\mu_E =0.4$). }
\label{Fig4}
%\end{minipage}
\end{figure*}
%%%%%%%%%%%%%%%%%%%%%%%%%%%%%%%%%%%%%%%%%%%%%%%%%%%%%%%%%%%%%%%%%%%%%%%%%%%%

The results presented above then motivate us to investigate how our bare quark (i.e. leading particle) suppression predictions, done with the dynamical energy loss, agree with the jet suppression measurements. To that end, in Fig.~\ref{Fig4}, we show our predictions of $R_{AA}$ {\it vs.} $p_\perp$ for the light (full curve), charm (dashed curve) and bottom (dot-dashed curve) probes. These leading particle predictions are shown together with inclusive jets from the ATLAS experiments~\cite{ATLAS_Jets} (left panel) and with both inclusive jets~\cite{CMS_Jets} and b-jets~\cite{CMS_BJets} from CMS (right panel). The predictions for both light and heavy probes are in a good agreement with the available jet measurements. This, together with the near overlap of the single particle and the jet suppression data shown in Fig. 5, therefore suggests that the leading particle predictions agree well with the jet $R_{AA}$ measurements.

There are few other important conclusions: {\it i}) above 50 GeV, we predict almost the same suppressions for the light, charm and bottom quarks (see also the middle panel of Fig.~\ref{RAA5TeV}); {\it ii})  this prediction, extrapolated from the single particle predictions to the light and b-jets, is in agreement with the measured experimental data. Since charm jet suppression is not yet measured, our result that the charm suppression overlaps with the light and bottom suppressions, likely suggests that c-jet $R_{AA}$ will overlap with both unidentified and b-jet $R_{AA}$s.

%%%%%%%%%%%%%%%%%%%%%%%% Fig. 5 %%%%%%%%%%%%%%%%%%%%%%%%%%%%%%%%%%%%%%%%%%%%
\begin{figure}
\epsfig{file=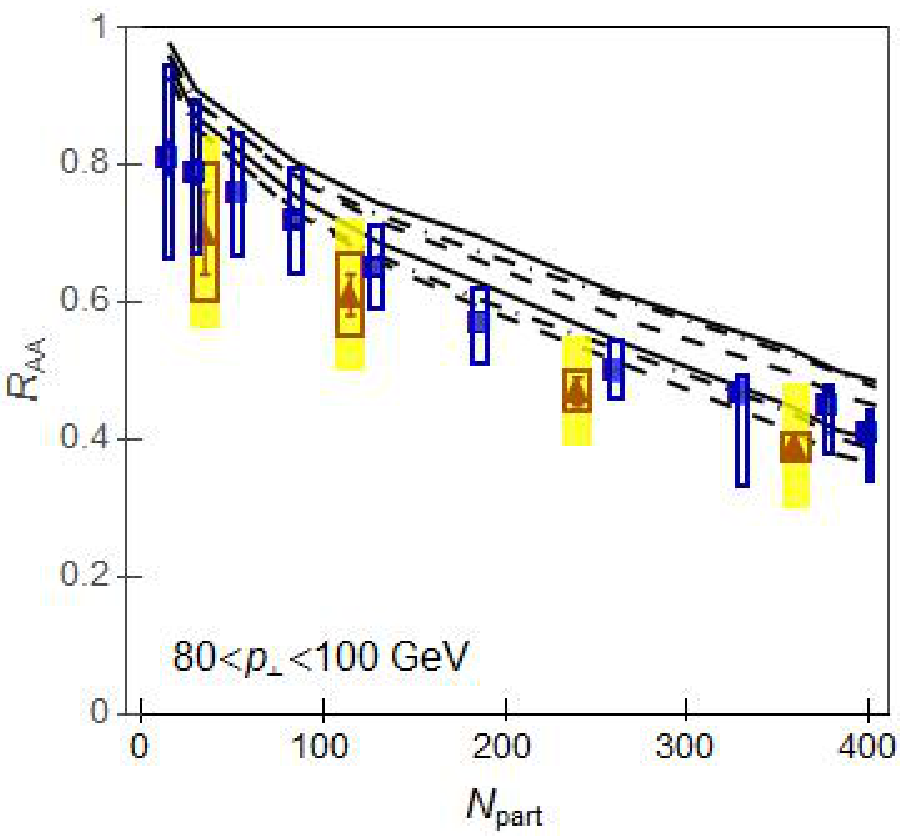,width=3in,height=3in,clip=5,angle=0}%\end{minipage}
\vspace*{-0.4cm}
\caption{{\bf Single particle suppression predictions {\it vs.} jet data.} Single particle predictions for $R_{AA}$ {\it vs.} $N_{part}$ are compared with ATLAS data for inclusive jets~\cite{ATLAS_Jets} (blue squares, $80<p_\perp<100$ GeV momentum region) and CMS data for b-jets~\cite{CMS_BJets} (orange triangles, $80<p_\perp<90$ GeV momentum region). White bands with dashed, dot-dashed and full boundaries, respectively, correspond to charm, bottom and light flavor predictions for $80<p_\perp<100$ GeV. The upper (lower) boundary of each band corresponds to $\mu_M/\mu_E =0.6$ ($\mu_M/\mu_E =0.4$). }
\label{Fig5}
%\end{minipage}
\end{figure}
%%%%%%%%%%%%%%%%%%%%%%%%%%%%%%%%%%%%%%%%%%%%%%%%%%%%%%%%%%%%%%%%%%%%%%%%%%%%

Finally, the similar conclusion is also obtained if the suppression is analyzed as a function of the number of participants (Fig. 7). In particular, we also see that $R_{AA}$ {\it vs.} $N_{part}$ single particle predictions for all three types of probes nearly overlap with each other and explain well the inclusive and b-jets data, which are also shown in the figure. Finally, the overlap of the suppression predictions is also consistent with the overlap in the data - similarly, as shown in Fig. 4, the  case of the charm jets is a new prediction to be tested by the future measurements.

\section{Conclusion}
In this paper, we systematically explored the mass tomography effects, which can be observed for different probes and in the wide momentum range corresponding to the span of the available experimental data. The predictions of the suppression dependence from both the momentum and the number of participants were generated and compared with the available single particle and jet measurements. As a result, we obtained both the agreement of the theoretical results with the available data, and generated new predictions to be tested in the upcoming experiments, as we briefly summarized below.

For the single particle predictions, we obtained that significant mass tomography effects can be noticed below 50 GeV, related with the difference between the bottom and the charm/light suppressions. While this difference is sometimes attributed to different initial distributions for the charm and bottom quarks, we here showed that this effect is almost entirely a consequence of the differences in the respective energy losses (i.e. the dead-cone effect), while the initial distribution contribution to the difference is almost negligible.

Furthermore, at the existing 2.76 TeV collision energy, we showed that the leading particle predictions agree well with the jet measurements.  Moreover, the experimental results show that there is a reasonable overlap between the single particle and jet suppression experimental data. These findings are interesting, particularly since our suppression approach does not include the features such as jet reconstruction~\cite{FastJet} (which are considered crucial for accurate description of (di)jet suppression~\cite{Jet1,Jet2,Jet3,Jet4,Jet5,Jet6,Jet7,Jet8}), but includes an advanced dynamical energy loss description for the leading parton.  Therefore, the agreement between the single particle and jet $R_{AA}$ measurements, both with respect to the experimental data and the theoretical predictions is currently unclear, and even if this agreement turns out to be accidental, understanding it may provide an important outlook for the future research.

Finally, we here provide clear predictions for the upcoming experimental data at 5.02 TeV collision energy: for the single particle data, we predict that, at the high momentum range $p_\perp > 100$ GeV, B and D meson (and likely c and b-jet) $R_{AA}$ data will nearly overlap with each other. On the other hand, our predictions for $h^\pm$ $R_{AA}$ unintuitively suggests a tendency for lower suppression compared to heavy mesons. We, however, show that this lower suppression is a pure consequence of fragmentation function effect on $h^\pm$, while finite mass effect is negligible in this region. Finally, we predicted significant mass tomography effects related with B meson suppression below 50 GeV. As discussed above, these predictions also provide specific guidelines on where future experimental efforts related to this goal should be concentrated. For example, given these results, we think that it is clearly beneficial to concentrate further efforts on improving b probe data in the relevant momentum region; this can include both directly measuring B mesons instead of non-prompt $J/\Psi$, reducing the uncertainties, as well increasing the number of available measurements for this important probe. With regards to this, note that the CMS experiment already published their measurement of the nuclear modification factor for fully reconstructed B mesons in p+Pb collisions~\cite{CMS_BMesons}, while such measurements in Pb+Pb collisions are expected to become available soon from ALICE.

{\em Acknowledgments:}
We thank Yen-Jie Lee for useful discussions. This work is supported by Marie Curie IRG within the $7^{th}$ EC Framework Programme (PIRG08-GA-2010-276913) and by the Ministry of Science of the Republic of Serbia
under project numbers ON173052 and ON171004.

\end{document}